\documentclass[%
reprint,
 amsmath,amssymb,
 aps,
floatfix,
]{revtex4-2}

\usepackage[encapsulated]{CJK}
\usepackage[linesnumbered,ruled,vlined]{algorithm2e}
\SetKwInput{KwInput}{Input}                
\SetKwInput{KwOutput}{Output}              
\usepackage{braket}
\usepackage{graphicx}
\usepackage{dcolumn}
\usepackage{bm}
\usepackage{hyperref}

\newtheorem{thm}{Theorem}
\newtheorem{prop}{Proposition}

\begin{document}
\begin{CJK*}{UTF8}{gbsn}
\preprint{APS/123-QED}

\title{Non-IID Quantum Federated Learning with One-shot Communication Complexity}

\author{Haimeng Zhao (赵海萌)}
\email{haimengzhao@icloud.com}
\affiliation{Zhili College, Tsinghua University, Beijing 100084, China}




\date{\today}

\begin{abstract}
Federated learning refers to the task of machine learning based on decentralized data from multiple clients with secured data privacy. Recent studies show that quantum algorithms can be exploited to boost its performance. However, when the clients' data are not independent and identically distributed (IID), the performance of conventional federated algorithms is known to deteriorate. In this work, we explore the non-IID issue in quantum federated learning with both theoretical and numerical analysis. We further prove that a global quantum channel can be exactly decomposed into local channels trained by each client with the help of local density estimators. This observation leads to a general framework for quantum federated learning on non-IID data with one-shot communication complexity. Numerical simulations show that the proposed algorithm outperforms the conventional ones significantly under non-IID settings.
\end{abstract}

\maketitle
\end{CJK*}

\newcommand{\tr}{\text{Tr}}


\section{Introduction}

Recent advances in artificial intelligence \cite{goodfellow2016deep} and quantum computing \cite{nielsen2010quantum} have given birth to an emerging field of quantum machine learning \cite{biamonte2017quantum, das2019machine}. By leveraging quantum advantage in machine learning tasks, quantum machine learning, has demonstrated unprecedented power in solving a wide range of problems. Notable examples include solving linear equations \cite{harrow2009quantum}, quantum supervised learning \cite{lloyd2014quantum, schuld2019quantum, havlivcek2019supervised, rebentrost2014quantum}, and quantum generative learning \cite{gao2018quantum, lloyd2018quantum, liu2018differentiable}. This line of research mostly focuses on developing quantum algorithms that can showcase quantum speed-up in machine learning problems.

One of the most important ingredients in machine learning is the data. In real-world applications, data are often distributed among multiple clients and cannot be gathered into a single joint dataset for various reasons. For example, medical data from patients across multiple hospitals \cite{rieke2020future} or cyber-physical attack data from Internet of Things devices \cite{khraisat2021critical} are sensitive and private, and cannot be directly shared without anonymization. Obstacles in data transmission may also forbid us from constructing a joint dataset. For example, the data size might be too large and therefore the transmission is too expensive, or we need quantum data that suffer from decoherence and are hard to preserve or transmit. 

Machine learning algorithms with such decentralized data have been developed under the name of federated learning \cite{konevcny2016federated, mcmahan2017comm}. The conventional solution is the federated averaging algorithm \cite{mcmahan2017comm}, or \textit{FedAvg}, in which multiple clients jointly train a global model by sharing only the model parameters/updates while keeping their data private. Its quantum extensions, dubbed \textit{qFedAvg}, have been proposed to incorporate quantum features such as quantum speed-up and blind computing \cite{li2021quantum, xia2021quantumfed, chen2021federated, chehimi2022quantum, yun2022slimmable}.

However, the classical \textit{FedAvg} algorithm has three shortcomings already identified in the literature. Firstly, it suffers from the non-IID quagmire \cite{zhao2018federated, hsieh2020non}, i.e. its performance is proved to deteriorate when the local data of different clients are not independent and identically distributed (IID). But real data are often heterogeneous or even multi-modal for different clients. Secondly, the joint training involves gradient sharing, so data privacy is threatened by attacks based on gradient inversion \cite{zhu2019deep, geiping2020inverting}. Thirdly, it requires many rounds of communication, which is often the bottleneck of real-world applications. To reduce communication burdens, multiple one-shot alternatives have been proposed \cite{guha2019one, salehkaleybar2021one, zhou2020distilled, kasturi2020fusion}.

{In this work, we discuss whether the non-IID quagmire exists in quantum federated learning. The answer is yes, and we support it with both theoretical analysis and numerical experiments. Then we move on to propose a solution to this problem. We prove that a global quantum channel can be exactly decomposed into channels trained by each client with the help of local density estimators. It provides a general framework, dubbed \textit{qFedInf}, for quantum federated learning on non-IID data. Meanwhile, \textit{qFedInf} is one-shot in terms of communication complexity and is free from gradient sharing. We further identify its connection to mixture of experts (MoE) and ensemble learning. Numerical experiments in highly non-IID settings demonstrate that the proposed framework outperforms the conventional algorithm significantly with only one communication round.}

\section{Main Results}

\subsection{Decentralized Quantum Data} \label{sec:data}
We begin by setting up a typical decentralized quantum dataset. Classical datasets can be regarded as a special case of it, where the data samples are orthogonal to each other. 

{In federated learning tasks, we are given a set of clients $\{C_i\}_{i=1}^n$, and we can use a set of orthonormal basis elements $\{\ket{C_i}\in\mathcal{H}_{C}\}_{i=1}^n$ to denote them.} Each of the clients has access to its own dataset containing $N_{C_i}$ samples $D_i = \{\ket{\psi_j^{C_i}}\}_{j=1}^{N_{C_i}}$ from the data Hilbert space $\mathcal{H}_{x}$. The dataset can be statistically represented by the density matrix $\rho_x^{C_i}=N_{C_i}^{-1}\sum_{j}\ket{\psi_j^{C_i}}\bra{\psi_j^{C_i}}$. 
Note that the density matrices of different clients may vary dramatically (i.e. non-IID). For example, in a multi-class classification problem, each client may have only seen samples from two or three classes.

{Since the construction of entanglement between macroscopic objects from distant places is in general very difficult and is not expected to be realized in the near future \cite{nielsen2010quantum}, we focus on the situation where there is no entanglement among the clients.}
Then the joint density matrix of both the data and the clients is represented by $\rho = \sum_{i=1}^n\rho_{x}^{C_i} \otimes p_{C_i} \ket{C_i}\bra{C_i}$, where $p_{C_i}=N_{C_i}/N$ is the statistical weight of client $C_i$, and $N=\sum_i N_{C_i}$ is the total number of samples. 
{For a classical dataset, the joint density matrix $\rho$ is diagonal, and its matrix elements correspond to the joint probability distribution $p(x, C_i)$. }

\begin{figure}
    \centering
    \includegraphics[width=\linewidth]{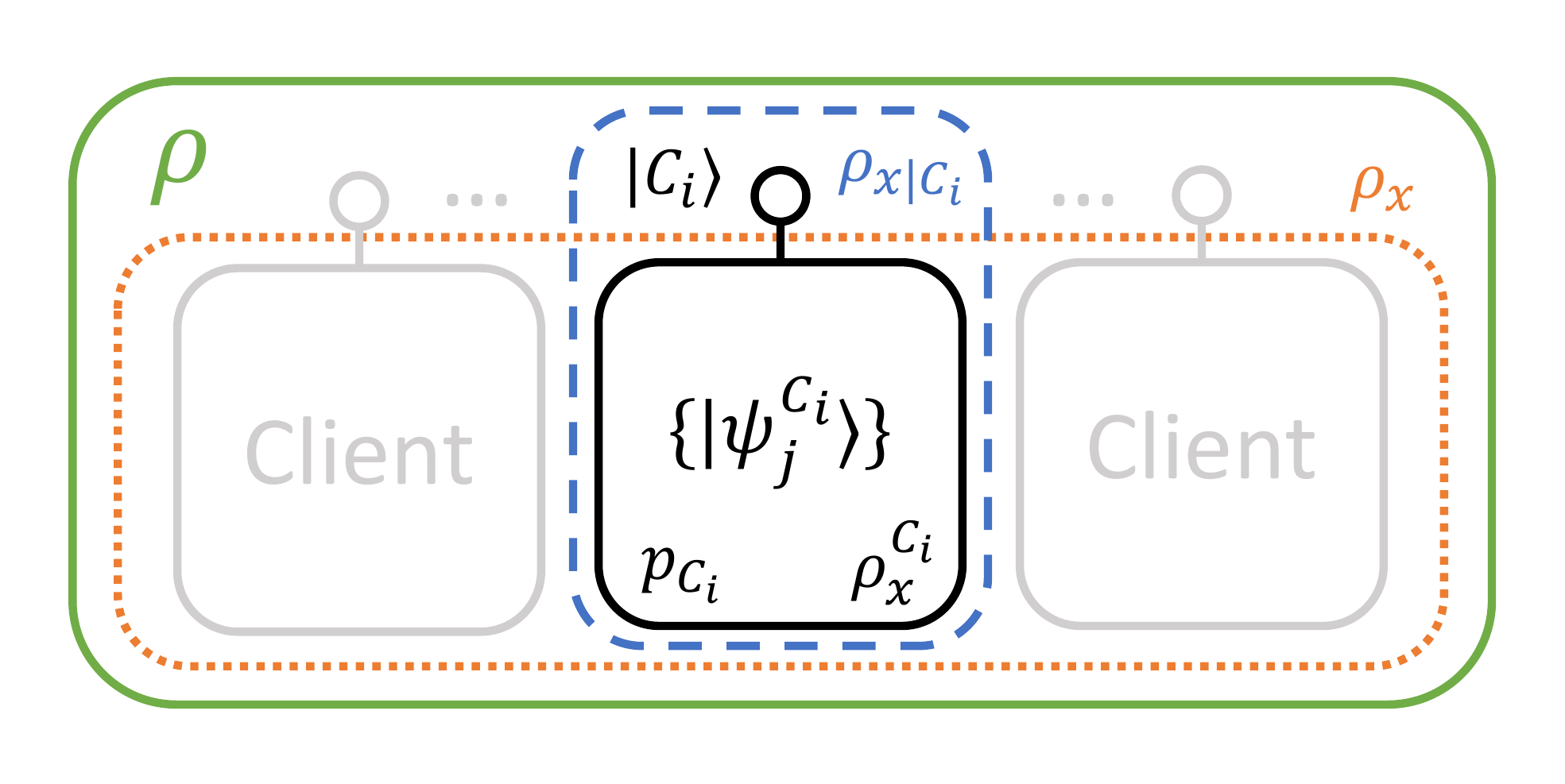}
    \caption{A schematic view of decentralized quantum data.}
    \label{fig:data}
\end{figure}

The averaged data density matrix is obtained by tracing out the clients: $\rho_x=\tr_{C}(\rho)=\sum_{i}p_{C_i}\rho_x^{C_i}.$ Let $P_{C_i}=\ket{C_i}\bra{C_i}$ be the projector onto the subspace of client $C_i$. Then we can introduce the conditional density matrix \cite{swart2020introduction, gonzalez2021classification} $\rho_{x|C_i} = {P_{C_i}\rho P_{C_i}}/{\tr(\rho P_{C_i})}=\rho_x^{C_i}\otimes \ket{C_i}\bra{C_i}$, which characterizes the data of a given client $C_i$. Tracing out the clients recovers the local dataset: $\tr_C(\rho_{x|C_i})=\rho_x^{C_i}$.
We summarize these concepts in Figure \ref{fig:data}.
{For supervised learning tasks such as classification, there will also be a label $y$ associated with each data sample $\ket{\psi}$. We have omitted the labels here for simplicity.}

\subsection{Quantum Federated Averaging}
Here we briefly review the quantum version of \textit{FedAvg} \cite{mcmahan2017comm}. 
In a general supervised quantum machine learning problem, we aim to find a quantum channel $\mathcal{M}$ that takes an input state $\ket{\psi}$ and transform it into the desired result (e.g. which class the image belongs to). We achieve this by tuning the variational parameters $w$ of the channel $\mathcal{M}_w$ to minimize the average of some loss function $\mathcal{L}$ on a given training dataset $D$:
\begin{equation} \label{eqn:goal}
    \min_{w} \frac{1}{N}\sum_{(\ket{\psi}, y)\in D} \mathcal{L}(w, \ket{\psi}, y),
\end{equation}
where $\ket{\psi}$ is the input and $y$ is the corresponding label.
We use gradient descent with learning rate $\eta$ to iteratively solve this optimization problem: at time step $t$,
\begin{equation} \label{eqn:update_cent}
    w_t = w_{t-1} - \eta \nabla_w \frac{1}{N}\sum_{(\ket{\psi}, y)\in D} \mathcal{L}(w, \ket{\psi}, y).
\end{equation}

For decentralized data, the total dataset is divided into local datasets from multiple clients: $D = \cup_i D_i$. Then we can decompose the update rule of Equation (\ref{eqn:update_cent}) into three steps: (1) local updates at time step $t$ for each client $C_i$,
\begin{equation} \label{eqn:update_local}
    w_t^i = w_{t-1}^i - \eta \nabla_w \frac{1}{N_{C_i}}\sum_{(\ket{\psi}, y)\in D_i} \mathcal{L}(w, \ket{\psi}, y),
\end{equation}
(2) global average for every $T$ steps, e.g. at time step $mT, m\in \mathbb{N}$,
\begin{equation} \label{eqn:update_avg}
    w_{mT} = \sum_{i=1}^n p_{C_i} w_{mT}^i,
\end{equation}
and (3) broadcast the averaged weights to all clients as the initial parameters for the next iteration. This is the basic protocol of \textit{qFedAvg}, the common ground of all the existing quantum federated learning algorithms \cite{li2021quantum, xia2021quantumfed, chen2021federated, chehimi2022quantum, yun2022slimmable}. It recovers the centralized update rule when $T=1$. However, in practice, this cannot be achieved since we use mini-batch training strategies.

\subsection{The Non-IID Quagmire of Quantum \textit{FedAvg}} \label{sec:noniid_th}
As pointed out in \cite{zhao2018federated, hsieh2020non}, \textit{FedAvg} faces the problem of non-IID quagmire, in which its performance deteriorates when the data of different clients are non-IID. Does this phenomenon also exist in the quantum regime? 

We follow the steps of \cite{zhao2018federated} and quantify the performance difference of \textit{FedAvg} and the centralized case by the weight divergence $\Delta = \| w^{(f)} - w^{(c)} \|$, where $w^{(f)}$ and $w^{(c)}$ are the weights given by \textit{qFedAvg} and the centralized case respectively. On the other hand, the level of non-IID can be quantified by the earth mover's distance (EMD) \cite{rubner2000earth} between the label distribution of client $C_i$,  $p^{(i)}(y=k)=\sum_{y\in D_i} \delta_{y, k}/N_{C_i}$, and the centralized distribution $p(y=k)=\sum_{y\in D} \delta_{y, k}/N_{C}$: $\text{EMD}_i = \sum_{k}\|p^{(i)}(y=k)-p(y=k)\|$. Then, as a direct quantum extension to Proposition 3.1 in \cite{zhao2018federated}, we have the following proposition:
\begin{prop} \label{prop:1}
For a loss function of the form $\mathcal{L}(w, \ket{\psi}, y) = \sum_{k=1}^{n_c} \delta_{y, k} f_k (w, \ket{\psi})$
\footnote{In a typical classification problem, $k$ denotes the different classes and the standard cross entropy loss is given by $f_k(w, \ket{\psi}) = \log r_k (w, \ket{\psi})$, where $r_k$ is the predicted probability of $\ket{\psi}$ belonging to class $k$.}, 
if the gradient $\nabla_w \mathbb{E}_{\ket{\psi}|y=k}f_k(w, \ket{\psi})$ defined as $\nabla_w (\sum_{\ket{\psi}, y} \delta_{y, k} f_k(w, \ket{\psi})/\sum_{\ket{\psi}, y}\delta_{y,k})$ is $\lambda_k$-Lipschitz for all possible value $k$ of label $y$, then the following inequality holds for the weight divergence of \textit{qFedAvg}:
\begin{equation}
\begin{split}
    \Delta_{mT} \le &\sum_{i=1}^n p_{C_i} (a_i)^T \Delta_{(m-1)T}\\
    &+ \eta \sum_{i=1}^n p_{C_i} \text{EMD}_i \sum_{j=1}^{T-1} (a_i)^j g(w^{(c)}_{mT-1-j}),
\end{split}
\end{equation}
where $g(w) = \max_k \| \nabla_w \sum_{\ket{\psi}, y} \delta_{y, k} f_k(w, \ket{\psi}) \|$ and $a_i = 1 + \eta \sum_k p^{(i)}(y=k)\lambda_{k}$.
\end{prop}
Therefore, EMD is indeed a source of the weight divergence when $T>1$. {This provides a theoretical explanation for the existence of non-IID quagmire in quantum federated learning, similar to its classical counterpart \cite{zhao2018federated}. We provide a detailed proof of Proposition \ref{prop:1} in Appendix \ref{app:prop}.}
Numerical experiments are conducted in Section \ref{sec:noniid} to empirically illustrate this phenomenon in quantum federated learning.

\subsection{Federated Decomposition of Quantum Channels}\label{sec:decomp}

Now that we have identified the non-IID quagmire in \textit{qFedAvg}, we aim to find a different approach to tackle Equation (\ref{eqn:goal}) when the data are decentralized. We consider the case where the loss function is a function of the output of the quantum channel and the label: $\mathcal{L}(w, \ket{\psi}, y) = \mathcal{L}(\mathcal{M}_w(\sigma_x^\psi), y)$, where $\sigma_x^\psi = \ket{\psi}\bra{\psi}$.

We begin by noting that when the data are decentralized, each client $C_i$ can still train a local channel $\mathcal{M}_i$ on its own data $D_i$:
\begin{equation} \label{eqn:local_goal}
    \mathcal{M}_i = \text{argmin}_\mathcal{M} \frac{1}{N_{C_i}}\sum_{(\ket{\psi}, y)\in D_i} \mathcal{L}(\mathcal{M}(\sigma_x^\psi), y).
\end{equation}
{To make use of the local channels, we can decompose the original problem, Equation (\ref{eqn:goal}), as}
\begin{equation} 
\begin{split}
    &\min_{\mathcal{M}} \frac{1}{N}\sum_{(\ket{\psi}, y)\in D} \mathcal{L}(\mathcal{M}(\sigma_x^\psi), y) \\
    &= \min_{\mathcal{M}} \sum_i p_{C_i} \left[ \frac{1}{N_{C_i}}\sum_{(\ket{\psi}, y)\in D_i} \mathcal{L}(\mathcal{M}(\sigma_x^\psi), y)\right]
\end{split}.
\end{equation}
{We can solve the whole minimization problem if we can solve each individual sub-problems simultaneously, which is exactly what $\mathcal{M}_i$ does.} Therefore, if we can construct a global channel $\mathcal{M}$, such that it coincides with all the local channels $\mathcal{M}_i$ on their own data, i.e.
\begin{equation}
    \mathcal{M}(\sigma_x^\psi) = \mathcal{M}_i(\sigma_x^\psi), \quad \forall \ket{\psi} \text{ from } D_i,
\end{equation}
then the goal is achieved. 
\footnote{The fact that identical samples have identical labels guarantees that the global channel $\mathcal{M}$ is well defined. For example, suppose we have two identical samples $\ket{\psi^{C_i}}=\ket{\psi^{C_j}}=\ket{\psi}$ with the same label $y$, but are from different clients $C_i\neq C_j$. Then, in order to fulfill the local minimization problems, Equation (\ref{eqn:local_goal}), we must have $\mathcal{M}_i(\ket{\psi}\bra{\psi})=\mathcal{M}_j(\ket{\psi}\bra{\psi})$.  }
{This can be rephrased backwards, as demanding that the local channels $\mathcal{M}_i$ coincide with the global channel $\mathcal{M}$ on condition that the input data $\sigma_x^\psi$ comes from its own dataset $D_i$, which is statistically represented by $\rho_x^{C_i}$.} That is, \begin{equation}
    \mathcal{M}_i(\sigma_x^\psi) 
    =\mathcal{M} \left(\frac{P_{x}^\psi\rho_{x}^{C_i} P_{x}^\psi}{\tr(\rho_{x}^{C_i} P_{x}^\psi)} \right)
    =\mathcal{M} \left(\tr_{C} \left(\frac{P_{x}^\psi\rho_{x|C_i} P_{x}^\psi}{\tr(\rho_{x|C_i} P_{x}^\psi)} \right) \right),
\end{equation}
where $P_x^\psi=\ket{\psi}\bra{\psi}$ denotes the projector.
\footnote{{To avoid confusion, we insist on using different notations for $P_{x}^\psi = \sigma_x^\psi = \ket{\psi}\bra{\psi}$ to emphasize their differences in physical meaning: we use $\sigma_x^\psi$ to denote the quantum state that you can load into your circuit, while we use $P_x^\psi$ to denote the projection operator.}}

To combine these local channels into a global one, we also need to capture the information of the local datasets $\rho_x^{C_i}$. To achieve this, we introduce the local density estimator $\mathcal{D}_i$, which is trained to output the probability density of the input state within the local dataset:
\begin{equation}
    \mathcal{D}_i(\sigma_x) = \tr(\rho_{x}^{C_i} P_x^\psi) = \tr(\rho_{x|C_i} P_x^\psi).
\end{equation}

In fact, the local channels $\mathcal{M}_i$ and density estimators $\mathcal{D}_i$ are enough to give an explicit construction of the global channel $\mathcal{M}$. This is provided by the following theorem (proved in Appendix \ref{app:proof}):

\begin{thm}\label{thm:1}~\\
(Federated Decomposition of Quantum Channels)\\
For each client $C_i, i=1, \cdots, n$, which only has access to its own data $\rho_x^{C_i}$ with $N_{C_i}$ samples, a local channel $\mathcal{M}_i$ and a local density estimator $\mathcal{D}_i$  can be trained. Assuming that there is no entanglement among clients, then the global channel $\mathcal{M}$ can be decomposed into
\begin{align}
    \mathcal{M}(\sigma_x) = \sum_{i=1}^n\mathcal{M}_i(\sigma_x) q_i = \sum_{i=1}^n\mathcal{M}_i(\sigma_x) \frac{\mathcal{D}_i(\sigma_x)p_{C_i}}{\sum_j \mathcal{D}_j(\sigma_x) p_{C_j}},
    \label{eqn:decomp}
\end{align}
where $q_i=\mathcal{D}_i(\sigma_x)p_{C_i}/\sum_j \mathcal{D}_j(\sigma_x) p_{C_j}$ is the decomposition weight, $\sigma_x$ is any pure input state and $p_{C_i} = N_{C_i}/\sum_{j=1}^n N_{C_j}$. 
Extension to mixed input states follows from direct linear superposition.
\end{thm}
We note that the classical special case of this theorem was first introduced in \cite{liu2022federated}.
As a result, for any input state $\sigma_x$, if we randomly apply a local channel $\mathcal{M}_i$ with probability $q_i$, the result will be statistically the same as the global channel $\mathcal{M}$. This fact leads to the following framework for quantum federated learning.

\subsection{A One-shot Quantum Federated Learning Framework for Non-IID Data}
\label{sec:framework}
Theorem \ref{thm:1} provides a framework for quantum federated learning. The specific protocol goes as follows. Firstly, each client $C_i$ trains a local channel $\mathcal{M}_i$ and a local density estimator $\mathcal{D}_i$ with its own data $\rho_x^{C_i}$. This step is completely distributed and concludes the whole training phase. Secondly, the trained channels $\mathcal{M}_i$ and density estimators $\mathcal{D}_i$ are sent to the server for inference according to Equation (\ref{eqn:decomp}). That is, when a new input $\sigma_x^\psi$ comes, the server computes $q_i$ via the density estimators. Then it randomly loads the parameters from $\mathcal{M}_i$ with probability $q_i$ and gathers the outcomes. We call this framework \textit{quantum federated inference}, or \textit{qFedInf} for short. The detailed algorithms are summarized in Algorithm \ref{alg:train} and \ref{alg:infer}.

\begin{algorithm}
\caption{Quantum Federated Inference: Training Phase on Client $C_i$}
\label{alg:train}
\KwInput{
The local data set $D_i$; number of training epochs $n_e$; batch size $n_b$; learning rate $\eta$; the optimizer $\text{Opt}$.
}
\KwOutput{
The trained channel $\mathcal{M}_i$ and density estimator $\mathcal{D}_i$ to be sent to the server.
}
Initialize the channel $\mathcal{M}_i$ with parameters $w$.\\
\For{$\mathrm{epoch} \gets 1$ \KwTo $n_e$}{
\For{$\mathrm{batch}~(\ket{\psi_j^{C_i}}, y_j)_{j=1}^{n_b} \in D_i$}{
Calculate the gradient $g \gets \nabla_w \sum_j \mathcal{L}(\mathcal{M}_{i, w}(\ket{\psi_j^{C_i}}\bra{\psi_j^{C_i}}), y_j)$.\\
Update the channel parameters $w \gets w - \eta\cdot\mathrm{Opt}(g)$.
}
}
Train the density estimator $\mathcal{D}_i$.\\
\Return{$\mathcal{M}_i, \mathcal{D}_i$.}
\end{algorithm}

\begin{algorithm}
\caption{Quantum Federated Inference: Inference Phase}
\label{alg:infer}
\KwInput{
A new sample $\ket{\psi}$ to perform inference on; channels $\{\mathcal{M}_i\}_{i=1}^n$ and density estimators $\{\mathcal{D}_i\}_{i=1}^n$ gathered from the clients.
}
\KwOutput{
The inference outcome $\mathcal{M}(\ket{\psi}\bra{\psi})$.
}
\For{$i \gets 1$ \KwTo $n$}{
Calculate the density $\tilde{q}_i \gets \mathcal{D}_i (\ket{\psi}\bra{\psi})$.\\
}
Calculate the weights $q_i \gets \tilde{q}_i p_{C_i} / \sum_{m} \tilde{q}_m p_{C_m}, \quad i=1, \cdots, n$.\\
Choose a random channel $\mathcal{M}_i$ with probability $q_i$.\\
$\mathcal{M}(\ket{\psi}\bra{\psi}) \gets \mathcal{M}_i (\ket{\psi}\bra{\psi})$.\\
\Return{$\mathcal{M}(\ket{\psi}\bra{\psi})$.}
\end{algorithm}

In practice, there is a wide range of choices for the channels $\mathcal{M}_i$ and the density estimators $\mathcal{D}_i$. For channels, we can use classical, quantum, or hybrid algorithms at our will \cite{bishop2006pattern, goodfellow2016deep, li2021recent}. 
{For density estimators, we can use classical ones like Gaussian mixture models \cite{bishop2006pattern} and normalizing flows \cite{rezende2015variational}, quantum-inspired ones like \cite{gonzalez2022learning}, or quantum ones such as classical shadow tomography \cite{huang2020predicting} and quantum state diagonalization algorithms \cite{larose2019variational, xin2021experimental}.} Each kind of density estimator comes with a different training strategy. We can also classify the possible scenarios based on the classical/quantum nature of the data and the channels:

\textbf{Classical Data \& Classical Channel:} 
This is a purely classical problem already discussed in \cite{liu2022federated}.

\textbf{Classical Data \& Quantum Channel:}
To apply quantum channels to classical data, one needs to choose an appropriate encoding scheme, e.g. amplitude encoding or gate encoding, to encode the data into quantum states. This gives rise to the problem of whether to use a classical density estimator on the original data or a quantum one on the encoded states. Note that before encoding, different samples are orthogonal to each other. However, the encoded quantum samples are in general overlapped, meaning that there's a possibility of mistaking one sample for another. Nevertheless, quantum density estimators offer a potential exponential speed-up. So there's a trade-off between accuracy and efficiency.

\textbf{Quantum Data \& Classical Channel:}
{In general, trying to apply classical channels to process quantum data is not very efficient, as the tomography and representation of a quantum state cost exponentially large classical computational resources. Nevertheless, proposals \cite{huang2022provably} have been made to use the classical shadows \cite{huang2020predicting} of a quantum state as the input to a classical machine learning algorithm. We defer to future works to investigate the performance of these proposals in a federated learning context.}

\textbf{Quantum Data \& Quantum Channel:}
For quantum data, we need a quantum density estimator to estimate $\tr(\rho_x^{C_i} P_x^\psi)$. This task can be solved by classical shadow tomography, which is proved to saturate the information-theoretic bound \cite{huang2020predicting}.

Compared to \textit{qFedAvg}, the proposed framework \textit{qFedInf} shares some merits with its classical counterpart \cite{liu2022federated}. It's one-shot in the sense that only one communication between the server and each client is required. Meanwhile, there's no need for a global public dataset or data synthesis/distillation, which are the common ground of existing one-shot algorithms \cite{zhou2020distilled, kasturi2020fusion}. Moreover, since no gradient information is transmitted, it's automatically immune to attacks based on gradient inversion \cite{zhu2019deep, geiping2020inverting}. Finally, the density estimators can capture possible data heterogeneity, providing a new way to perform federated learning with non-IID data.

{Though we have mainly focused on supervised learning throughout this paper, we also note that Theorem \ref{thm:1} holds for generic quantum channels, which is the most general form of quantum information processing. So the proposed framework \textit{qFedInf} may be applied to machine learning tasks beyond classification. In Appendix \ref{app:generative} we provide an example of applying \textit{qFedInf} to perform quantum generative learning \cite{gao2018quantum, lloyd2018quantum, liu2018differentiable}.}

{Finally, we give a brief discussion on the complexity of the proposed framework. Suppose that the number of clients is $N_\text{client}$, the number of iterations needed for training is $N_\text{iter}$, and the number of parameters in each of the quantum channels $\mathcal{M}_i$ and density estimators $\mathcal{D}_i$ are $N_\text{channel}$ and $N_\text{density}$. Then the communication complexity \cite{konevcny2016federated} of the proposed \textit{qFedInf} is $\mathcal{O}(N_\text{client} (N_\text{channel} + N_\text{density}))$. In comparison, the communication complexity of \textit{qFedAvg} is $\mathcal{O}(N_\text{client} N_\text{iter} N_\text{channel})$, which is much less efficient when $N_\text{iter}$ is large. As for the circuit complexity \cite{li2021multilevel, barenco1995elementary}, it depends on the specific choice of the quantum circuit used as the channels. If the channels have a quantum speed-up, so will \textit{qFedInf}.}

\subsection{Connection to MoE and Ensemble Learning} \label{sec:ensemble}
{
In this section, we discuss the connection between the proposed framework \textit{qFedInf} and mixture of experts (MoE) \cite{masoudnia2014mixture, jordan1994hierarchical, jacobs1991adaptive}, which is an important strategy in ensemble learning. The idea of ensemble learning is to combine several models to make a joint prediction \cite{zhou2021machine, sagi2018ensemble}. Different models are expected to compensate for each other, leading to better performance. MoE, as a special kind of ensemble learning method, consists of two parts: a set of functions serving as the judge that decides the relative weight of different models, called the gating function; and an ensemble of specialized models, called the experts, each expected to perform well only on a subset of the total input space. In practice, MoE, along with other ensemble approaches such as bagging and boosting, has given birth to many state-of-the-art solutions to a wide range of machine learning problems \cite{zhou2021machine, sagi2018ensemble, masoudnia2014mixture}.
}

{
In hindsight, we can see that \textit{qFedInf} also consists of two parts: a set of density estimators $\mathcal{D}_i$ that decides the probabilistic weight of different local channels $\mathcal{M}_i$, as in Equation (\ref{eqn:decomp}); and a set of local channels $\mathcal{M}_i$, each performing well only on the client's own data. In the language of MoE, the density estimators are the gating functions, and the local channels correspond to the experts. Therefore, \textit{qFedInf} is exactly an MoE working in the one-shot federated learning context. In Section \ref{sec:exp}, numerical experiments show that \textit{qFedInf} can indeed combine the knowledge of many weak classifiers (shallow circuits only capable of binary classification) to achieve the capability of large models.
}

\section{Numerical Experiments} \label{sec:exp}

\begin{figure}
    \centering
    \includegraphics[width=\linewidth]{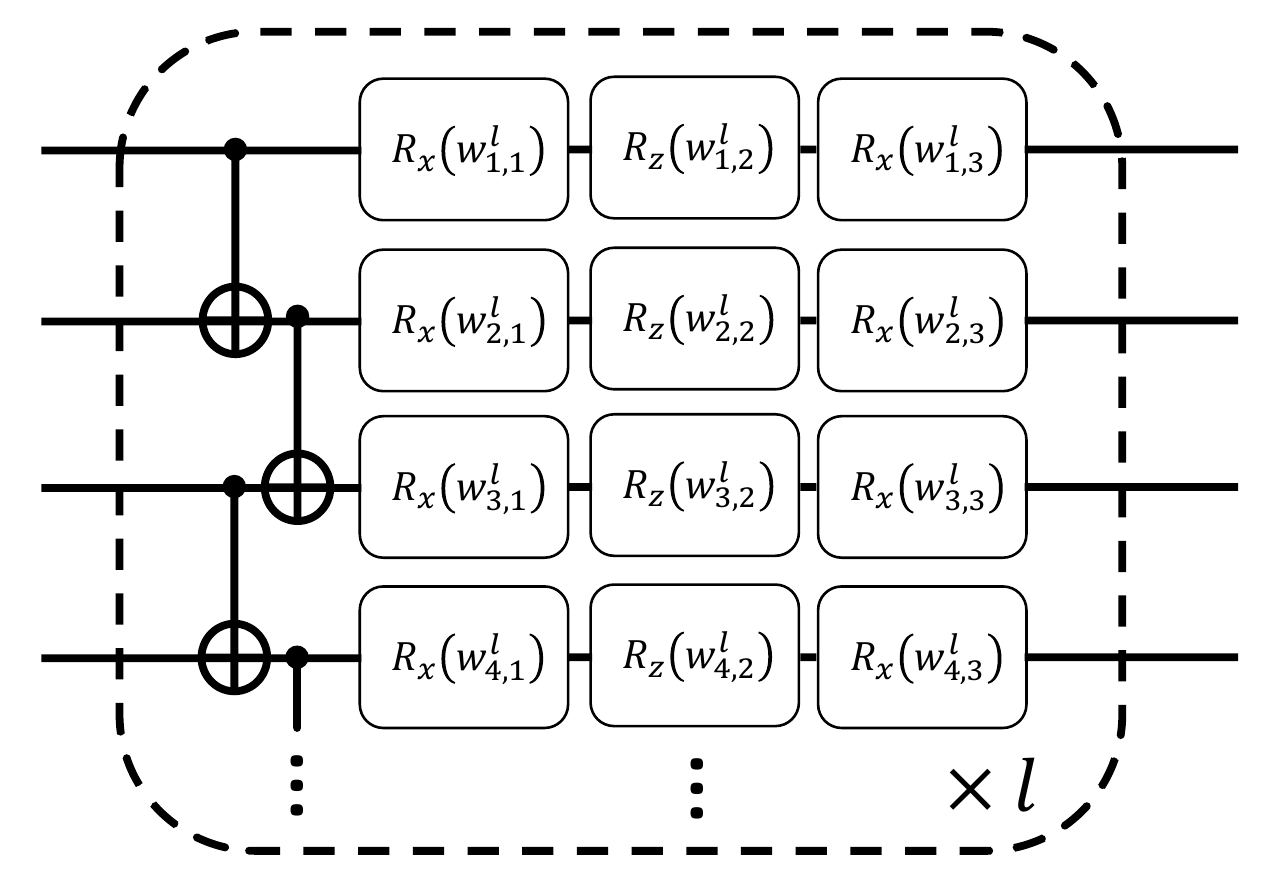}
    \caption{The quantum circuit used as the quantum classifier.}
    \label{fig:circuit}
\end{figure}

\begin{table*}
\centering
\begin{tabular}{c|ccccc}
\hline
Top-1 Accuracy/\% & Centralized & \textit{qFedInf}($\star$) & \textit{qFedAvg}($\star$) & \textit{qFedInf}($\circ$) & \textit{qFedAvg}($\circ$) \\
\hline
MNIST          & 91.2$\pm$0.6      & 92.4$\pm$0.3           & 86.2$\pm$1.0 & 92.7$\pm$0.2 & 88.4$\pm$0.8           \\
Fashion-MNIST  & 77.2$\pm$0.5      & 74.0$\pm$0.3           & 61.4$\pm$1.4 & 75.4$\pm$0.3 & 66.7$\pm$1.3           \\
\hline
\end{tabular}
\caption{Performance of centralized, \textit{qFedInf} and \textit{qFedAvg} classifiers on MNIST and Fashion-MNIST datasets under different federated settings ($\star$/$\circ$ for \textit{star}/\textit{cycle-2 structure}). The averages and standard deviations are obtained from 10 runs.}
\label{tab:perf}
\end{table*}

\begin{figure*}
\begin{minipage}{0.48\linewidth}
    \centering
    \includegraphics[width=\linewidth]{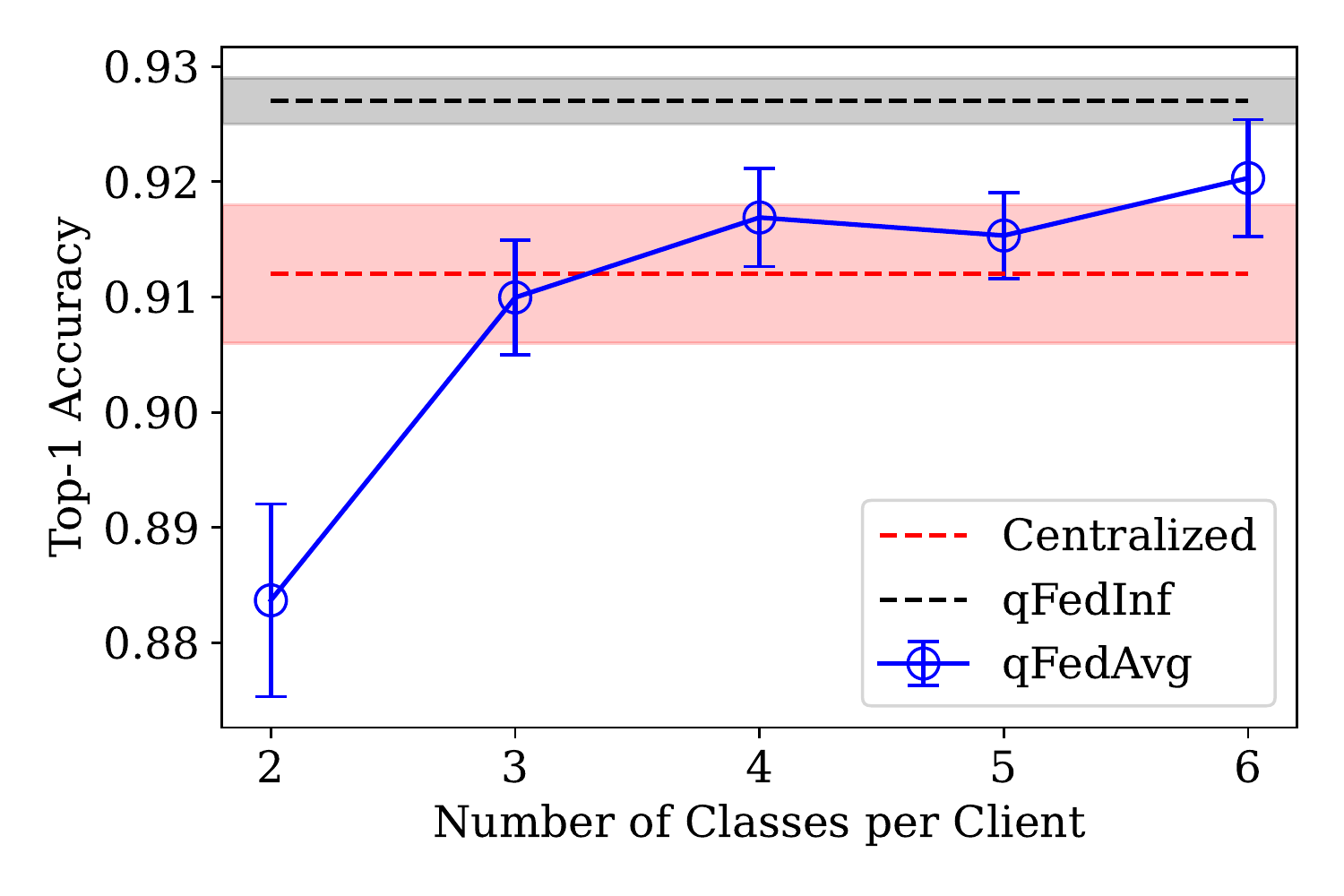}
\end{minipage}
\begin{minipage}{0.48\linewidth}
    \centering
    \includegraphics[width=\linewidth]{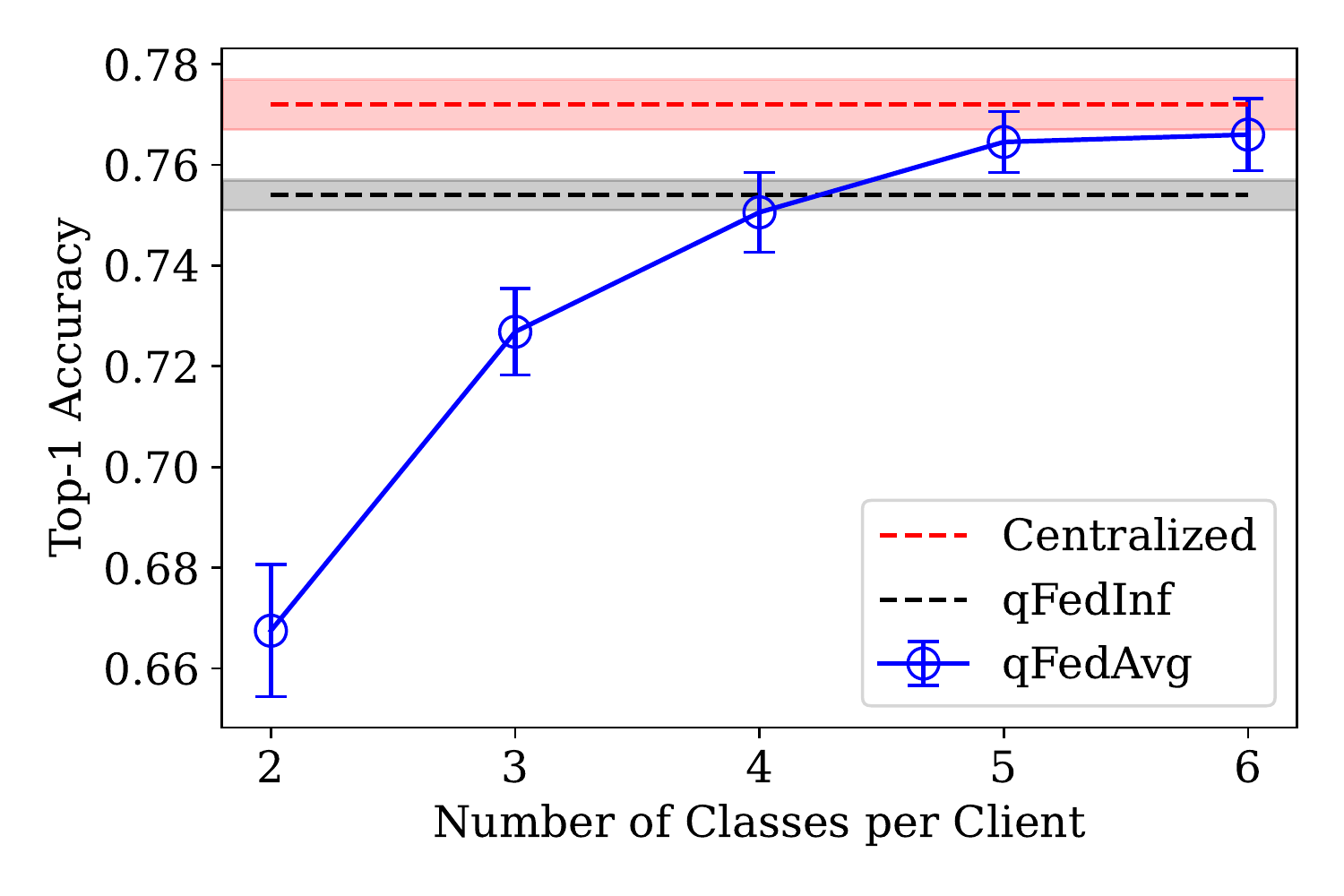}
\end{minipage}
\caption{The test accuracy of \textit{qFedAvg} on MNIST (left) and Fashion-MNIST (right) with different levels of non-IID in the \textit{cycle-m structure}. 
{The dashed line represents the test accuracy of centralized classifier (red) and \textit{qFedInf} (black) trained under the most heterogeneous setting (2 classes per client). The accuracy of \textit{qFedInf} is roughly unaffected by the number of classes per client.}
The error bars and shaded regions mark the standard deviation over 10 runs.}
\label{fig:noniid}
\end{figure*}

\subsection{Constructing Non-IID Datasets}
We observe that the existing quantum federated learning algorithms can already achieve high accuracies on binary classification tasks with synthetic and common classical/quantum datasets \cite{chehimi2022quantum, li2021quantum}. Therefore, to better illustrate the performance differences between \textit{qFedInf} and \textit{qFedAvg}, we devise a highly heterogeneous federated dataset based on 8 classes (``0'' through ``7'') from the MNIST handwritten digits dataset. To produce heterogeneity, we adopt the \textit{star structure} and \textit{cycle-m structure} settings from \cite{liu2022federated} as follows.

\textbf{\textit{Star structure:}} Each client only has access to the data of two classes, with one of the classes fixed. That is, client $C_i, i=1, \cdots, 7$ only has access to the data of digit ``0'' and ``$i$''.

\textbf{\textit{Cycle-m structure:}} Each client only has access to the data of $m$ classes in a cyclic way. Client $C_i, i=1, \cdots, 7$ has access to the data of digit ``$i$'', ..., ``$i+m-1$ module 8''.

Datasets of the same structures are also prepared for the Fashion-MNIST dataset, which is composed of images of daily objects and is regarded as a harder version of MNIST. 

\subsection{Details of the Quantum Classifiers}
We parameterize the quantum classifier as a quantum circuit of $l$ layers. Each layer contains a set of controlled-NOT gates on adjacent qubits, followed by a parameterized rotation on each qubit. The detailed circuit is shown in Figure \ref{fig:circuit}.

To load the data into an 8-qubit quantum circuit, we interpolate and resize each image into a size of 16 $\times$ 16 and use amplitude encoding to transform it into a quantum state \cite{biamonte2017quantum, li2014multidimensional}:
\begin{equation}
    \{x_j\}_{j=1}^{256} \rightarrow \ket{\psi} = \sum_{j=1}^{256} \frac{x_j}{\sqrt{\sum_{u=1}^{256}\|x_u\|^2}} \ket{j},
\end{equation}
where $\{x_j\}$ are the pixel values of the image and $\{\ket{j}\}$ denote the computational basis. 
{In experiments, amplitude encoding can be implemented using quantum random access memory \cite{giovannetti2008quantum, giovannetti2008architectures} or universal gate decomposition \cite{barenco1995elementary, long2001efficient, plesch2011quantum}.}

To perform inference, we measure the expectation values $\braket{Z_k}$ on each qubit, amplify the outcomes by a factor of 10, and fed them into the softmax function to predict the probability of each class. We use the Adam optimizer \cite{kingma2017adam} of learning rate $10^{-2}$ and a batch size of $128$ to minimize the standard cross entropy loss function. The gradients used in the optimization can be computed via the parameter shift rule \cite{mitarai2018quantum, li2021quantum, li2017hybrid}.

All the numerical simulations are conducted with JAX \cite{jax2018github} and TensorCircuit \cite{zhang2022tensorcircuit} on one NVIDIA Tesla V100 GPU. The source code is available at \url{https://github.com/JasonZHM/quantum-fed-infer}.

\subsection{Performance of \textit{qFedAvg} and \textit{qFedInf} on Non-IID Data} \label{sec:noniid}

\begin{figure*}
\begin{minipage}{0.48\linewidth}
    \centering
    \includegraphics[width=\linewidth]{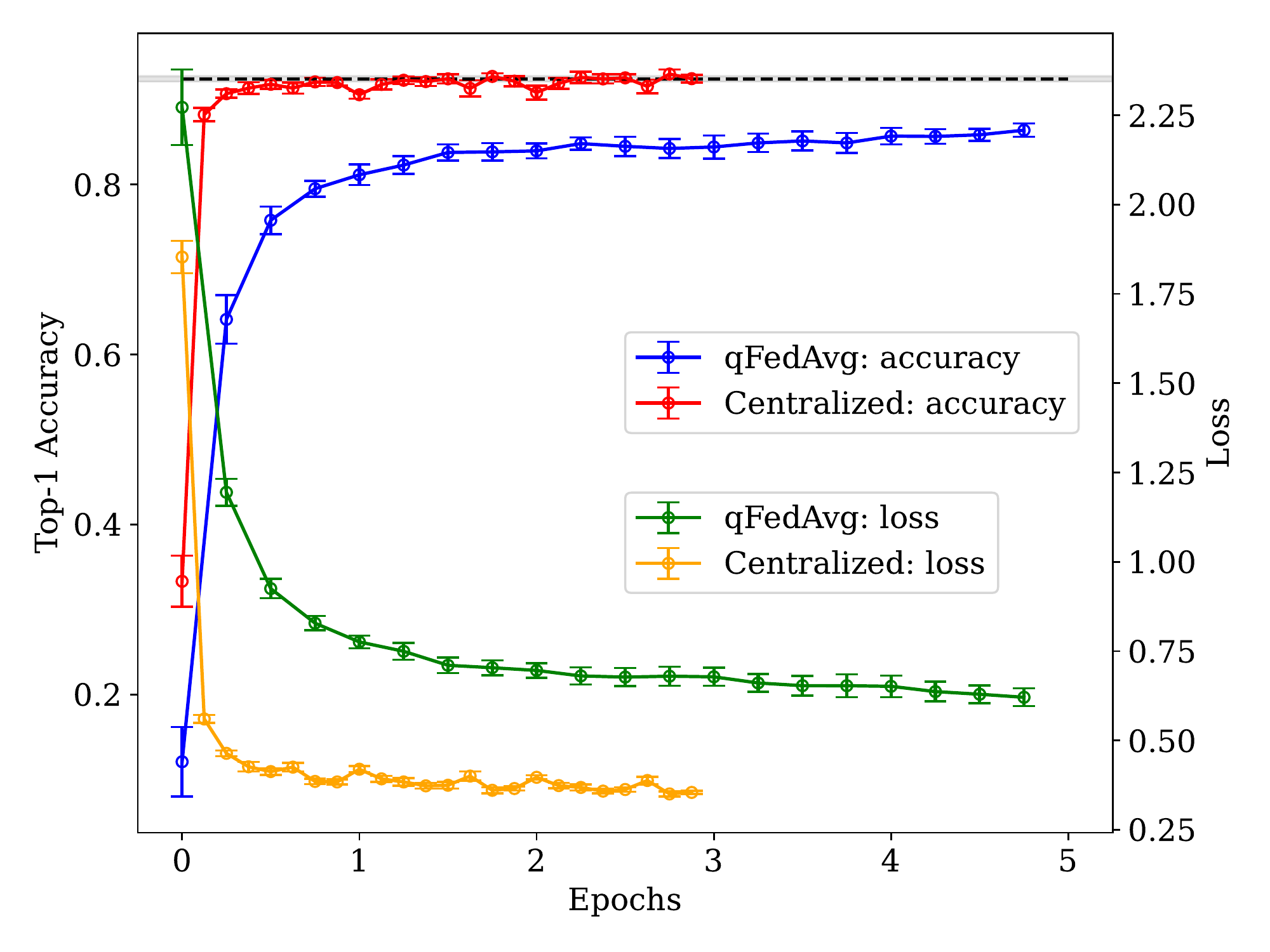}
\end{minipage}
\begin{minipage}{0.48\linewidth}
    \centering
    \includegraphics[width=\linewidth]{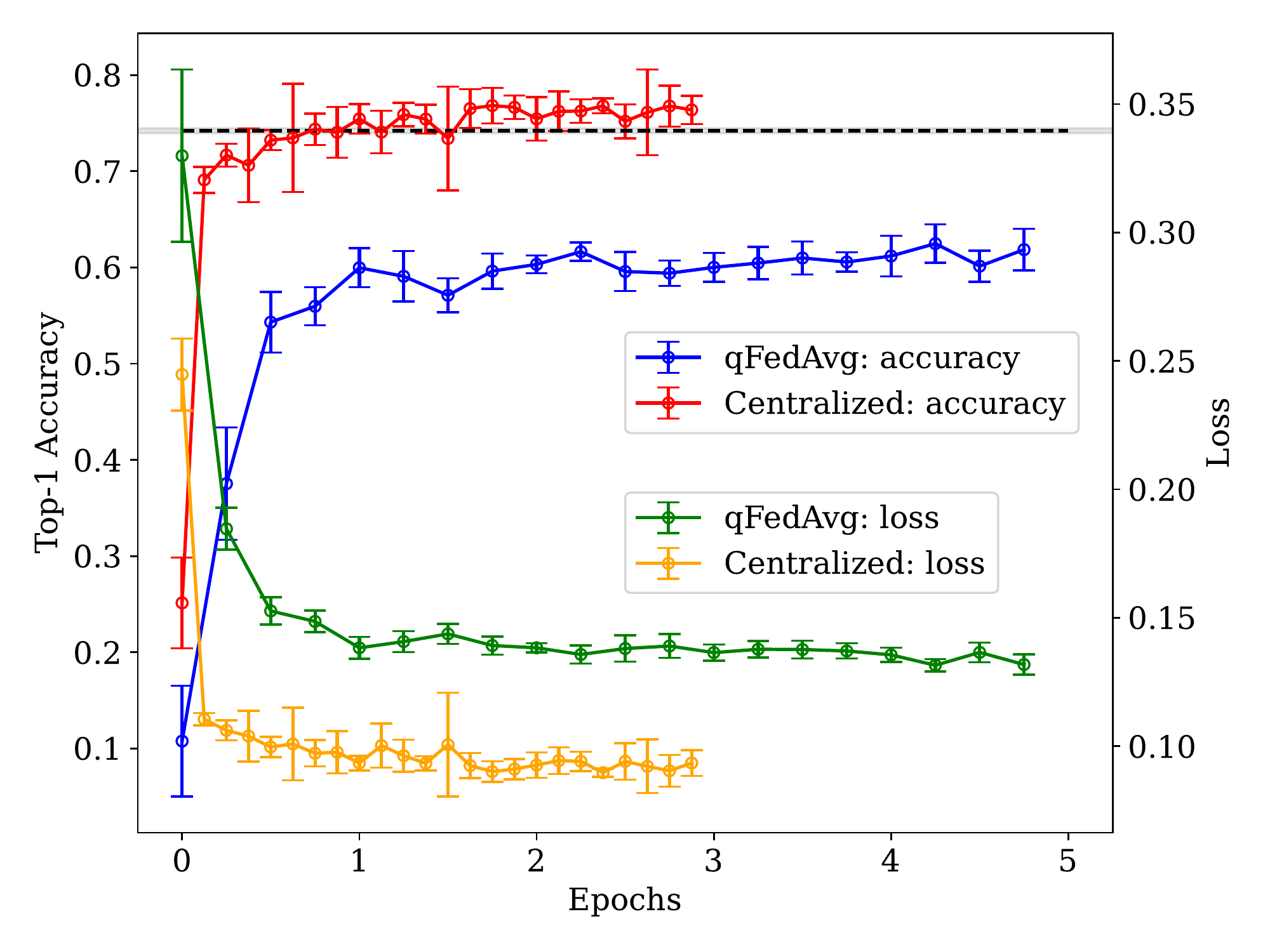}
\end{minipage}
\caption{The test loss and accuracy of centralized classifier and \textit{qFedAvg} on \textit{star structure} MNIST (left) and Fashion-MNIST (right). The dashed line represents the test accuracy of \textit{qFedInf}. The error bars and shaded regions mark the standard deviation over 10 runs.}
\label{fig:cent_fedavg}
\end{figure*}

We begin by demonstrating the non-IID quagmire of \textit{qFedAvg} discussed in Section \ref{sec:noniid_th} with numerical simulations. Specifically, we train and test \textit{qFedAvg} on datasets of the \textit{cycle-m structure}. The parameter $m$ serves as a good controller over the level of non-IID: as $m$ increases, each client will have access to more classes, and the level of non-IID will decrease.

We train the channel with $l=48$ and $m=2, 3, 4, 5, 6$ using \textit{qFedAvg} for 5 epochs. The global synchronization frequency of \textit{qFedAvg} is set to be one time per batch step. The resulting test accuracies on a test set of size 1024 are plotted in Figure \ref{fig:noniid}. In line with the theoretical analysis, we find that the top-1 accuracy increases as the level of non-IID drops. When the data are highly heterogeneous ($m=2$), \textit{qFedAvg} suffers from a loss of $\sim 4\%$ ($10\%$) in accuracy on MNIST (Fashion-MNIST) compared to the benchmark trained on the centralized data with the same circuit structure. Nevertheless, when the data heterogeneity is mild ($m \ge 5$), \textit{qFedAvg} can achieve comparable performance compared with the centralized classifier. We also plot the test loss and accuracy curves on \textit{star structure} in Figure \ref{fig:cent_fedavg}. As expected, \textit{qFedAvg} converges much slower to significantly lower accuracy.

\begin{figure*}
\begin{minipage}{0.48\linewidth}
    \centering
    \includegraphics[width=\linewidth]{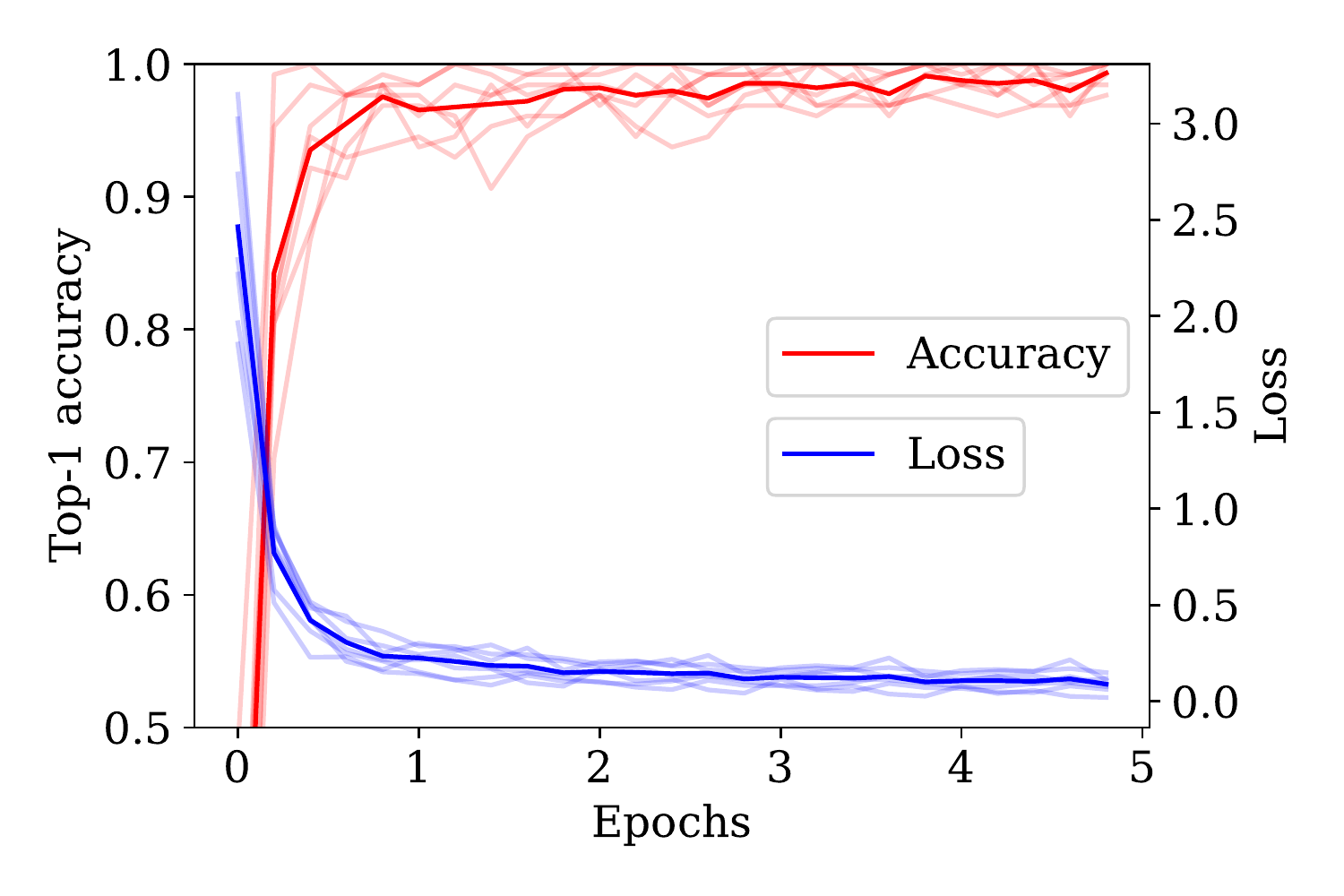}
\end{minipage}
\begin{minipage}{0.48\linewidth}
    \centering
    \includegraphics[width=\linewidth]{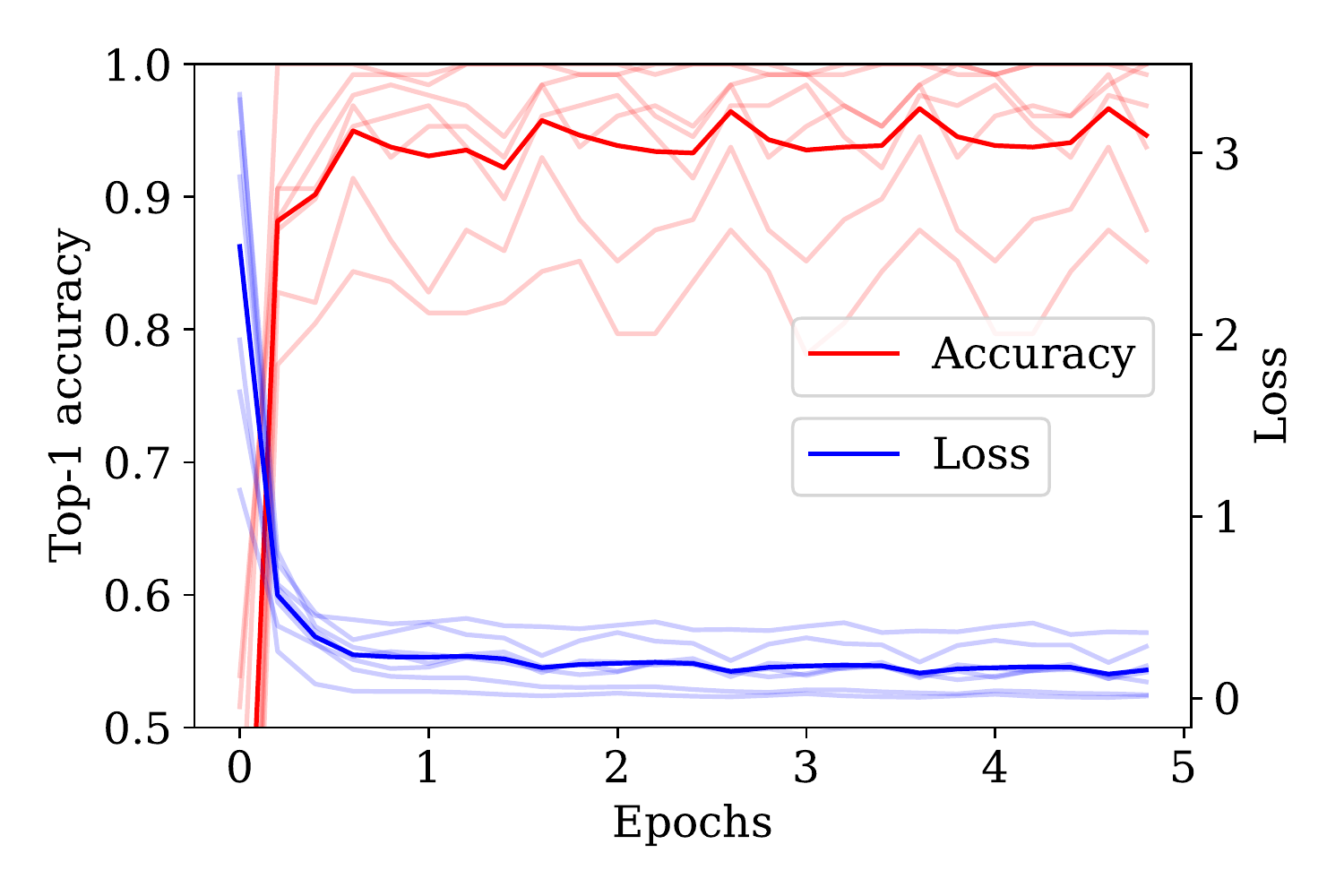}
\end{minipage}
\caption{The training loss and accuracy of the local classifiers $\mathcal{M}_i$ in \textit{qFedInf} on \textit{start structure} MNIST (left) and Fashion-MNIST (right). The average curves are highlighted.}
\label{fig:client}
\end{figure*}

To test the performance of \textit{qFedInf} on non-IID data, we train and test \textit{qFedInf} on the most heterogeneous settings, namely the \textit{star structure} and \textit{cycle-2 structure}, and compare it with \textit{qFedAvg} and the centralized benchmark. In such settings, each client's local classifier only needs to perform a binary classification, and thus in practice, we find that circuits with only a few layers suffice to achieve good performance. 
{For comparison with the $l=48$ \textit{qFedAvg} and the centralized benchmark, we choose $l=6$ for the local classifiers, so that the total number of variational parameters of the $8$ clients remains the same.} We train the local classifiers for 5 epochs and plot the loss and accuracy curves in Figure \ref{fig:client}. We adopt Gaussian mixture models with 5 modes as the local density estimators. The combined global model achieves a top-1 accuracy similar to the centralized benchmark and significantly higher than that of \textit{qFedAvg} on both settings of both datasets. The detailed accuracies are listed in Table \ref{tab:perf}.
{Meanwhile, the performance of \textit{qFedInf} is roughly unaffected by the number of classes per client, demonstrating its robustness against the level of non-IID.}

We note that in the training process, \textit{qFedAvg} requires a total number of 500 communication rounds, while \textit{qFedInf} only needs one. Moreover, the local classifiers used in \textit{qFedInf} are much shallower compared to those of \textit{qFedAvg} and the centralized benchmark. If we change the $l$ of the centralized classifier to 6, its test accuracy will drop to only $\sim 75\%$. 
{To rule out the possibility that barren plateau \cite{mcclean2018barren} causes this performance difference, we also test \textit{qFedInf} with $l=48$, and the resulting performance is the same as $l=6$, which suggests that the barren plateau issue is not significant in our settings.
Therefore, putting the federated settings aside, \textit{qFedInf} can indeed utilize the collective knowledge of many small models to achieve the capability of a large model, in line with MoE and ensemble learning as discussed in Section \ref{sec:ensemble}.}

\section{Conclusions}
In this work, we tackle the problem of non-IID data in quantum federated learning. We give a theoretical analysis of the non-IID quagmire in \textit{qFedAvg} and support it with numerical experiments. We prove that a global quantum channel can be exactly decomposed into local channels trained by each client with the help of local density estimators. It leads to a general framework \textit{qFedInf} for quantum federated learning on non-IID data. It's one-shot in terms of communication complexity and immune to attacks based on gradient inversion. 

We conduct numerical experiments on multi-class classification tasks to demonstrate the proposed framework. We devise a highly heterogeneous federated dataset based on MNIST and Fashion-MNIST. Experiments show that \textit{qFedInf} achieves a comparable performance compared to the centralized benchmark, and outperforms \textit{qFedAvg} with significantly fewer communication rounds. 

The non-IID issue has been regarded as a major challenge and is under active research in the literature of classical federated learning. Future works may focus on a more thorough analysis of this issue in the quantum regime: develop more challenging quantum federated datasets to demonstrate the quantum non-IID quagmire and test the performance of different density estimators. On the other hand, as quantum channels are the most general form of quantum information processing, we expect that more quantum machine learning algorithms can be made federated through the proposed framework. 
{Moreover, more quantum features such as quantum speed-up and universal blind quantum computation \cite{broadbent2009universal, li2021quantum} may also be incorporated into \textit{qFedInf} in the future.}

\begin{acknowledgments}
We thank Weikang Li, Jingyi Zhang, Yuxuan Yan, Rebing Wu, and Yuchen Guo for their insightful discussions.
We thank the anonymous reviewers for their constructive suggestions on the manuscript.
We also acknowledge the Tsinghua Astrophysics High-Performance Computing platform for providing computational and data storage resources. 
This work is financially supported by Zhili College, Tsinghua University.
\end{acknowledgments}
\begin{widetext}
\appendix

\section{Proof of Proposition \ref{prop:1}} \label{app:prop}
{Proposition \ref{prop:1} is a quantum generalization of its classical counterpart, Proposition 3.1 in \cite{zhao2018federated}. Below, we provide a detailed proof following the ideas introduced in \cite{zhao2018federated}.} Based on the definition of $\Delta$ and the update rules, Equations (\ref{eqn:update_cent}), (\ref{eqn:update_local}) and (\ref{eqn:update_avg}), we have

\begin{equation}
\begin{split}
    &\Delta_{mT} = \|w^{(f)}_{mT}-w^{(c)}_{mT}\| \\
    &= \|\sum_i p_{C_i} (w^i_{mT-1} - \eta \nabla_w \frac{1}{N_{C_i}}\sum_{(\ket{\psi}, y)\in D_i}\sum_{k=1}^{n_c}\delta_{y, k} f_k(w^i_{mT-1}, \ket{\psi})) -  w^{(c)}_{mT-1} + \eta \nabla_w \frac{1}{N}\sum_{(\ket{\psi}, y)\in D}\sum_{k=1}^{n_c}\delta_{y, k} f_k(w^{(c)}_{mT-1}, \ket{\psi})\| \\
    &= \|\sum_i p_{C_i} (w^i_{mT-1} - \eta \sum_{k=1}^{n_c}p^{(i)}(y=k) \nabla_w \mathbb{E}_{\ket{\psi}|y=k} f_k(w^i_{mT-1}, \ket{\psi})) - w^{(c)}_{mT-1} + \eta \sum_{k=1}^{n_c}p(y=k)\nabla_w \mathbb{E}_{\ket{\psi} | y=k} f_k(w^{(c)}_{mT-1}, \ket{\psi})\|. 
\end{split}
\end{equation}
Now we apply the triangle inequality and the Lipschitz conditions. Together with the definition $a_i = 1 +\eta \sum_k p^{(i)}(y=k)\lambda_k$, we have
\begin{equation} \label{eqn:A2}
\begin{split}
    \Delta_{mT} &\le \|\sum_i p_{C_i}w^i_{mT-1} - w^{(c)}_{mT-1}\| \\
    &+ \eta \| \sum_i p_{C_i} \sum_k  p^{(i)}(y=k) (\nabla_w \mathbb{E}_{\ket{\psi}|y=k} f_k(w^i_{mT-1}, \ket{\psi})) - \nabla_w \mathbb{E}_{\ket{\psi} | y=k} f_k(w^{(c)}_{mT-1}, \ket{\psi}))\| \\
    &\le \sum_i p_{C_i} (1+\eta\sum_k p^{(i)}(y=k)\lambda_k) \| w^i_{mT-1} - w^{(c)}_{mT-1} \| \\
    &= \sum_i p_{C_i} a_i \| w^i_{mT-1} - w^{(c)}_{mT-1} \|.
\end{split}
\end{equation}
Then we continue going backwards in the time steps. With the triangle inequality and the definitions of $g(w)$ and $\mathrm{EMD}_i$, we have
\begin{equation}
\begin{split}
    &\| w^i_{mT-1} - w^{(c)}_{mT-1} \| \\
    &\le \| w^i_{mT-2} - w^{(c)}_{mT-2}\| + \eta \| \sum_k  p^{(i)}(y=k) \nabla_w \mathbb{E}_{\ket{\psi}|y=k} f_k(w^i_{mT-2}, \ket{\psi})) - \sum_k  p(y=k) \nabla_w \mathbb{E}_{\ket{\psi}|y=k} f_k(w^{(c)}_{mT-2}, \ket{\psi}))\| \\
    &\le \| w^i_{mT-2} - w^{(c)}_{mT-2}\| \\
    &+ \eta \| \sum_k  p^{(i)}(y=k) \nabla_w \mathbb{E}_{\ket{\psi}|y=k} f_k(w^i_{mT-2}, \ket{\psi})) - \sum_k  p^{(i)}(y=k) \nabla_w \mathbb{E}_{\ket{\psi}|y=k} f_k(w^{(c)}_{mT-2}, \ket{\psi}))\| \\
    &+ \eta \|\sum_k (p^{(i)}(y=k)-p(y=k))\mathbb{E}_{\ket{\psi}|y=k} f_k(w^{(c)}_{mT-2}, \ket{\psi})) \| \\
    &\le a_i\| w^i_{mT-2} - w^{(c)}_{mT-2}\| + \eta g(w^{(c)}_{mT-2})\mathrm{EMD}_i.
\end{split}
\end{equation}
By induction and the broadcast rule $w^i_{(m-1)T} = w^{(f)}_{(m-1)T}$, we arrive at
\begin{equation}
\begin{split}
    \| w^i_{mT-1} - w^{(c)}_{mT-1} \| &\le a_i^{T-1}\| w^i_{(m-1)T} - w^{(c)}_{(m-1)T} \| + \eta \mathrm{EMD}_i \sum_{j=0}^{T-2}a_i^j g(w^{(c)}_{mT-2-j}) \\
    &= a_i^{T-1}\Delta_{(m-1)T} + \eta \mathrm{EMD}_i \sum_{j=0}^{T-2}a_i^j g(w^{(c)}_{mT-2-j})
\end{split}
\end{equation}
Plug it into Equation (\ref{eqn:A2}), and we finally reach the desired result:
\begin{equation}
\begin{split}
    \Delta_{mT} &\le \sum_i p_{C_i} a_i^T \Delta_{(m-1)T} \\
    &+ \eta \sum_i p_{C_i} \mathrm{EMD}_i \sum_{j=1}^{T-1}a_i^j g(w^{(c)}_{mT-1-j}).
\end{split}
\end{equation}

\section{Proof of Theorem \ref{thm:1}} \label{app:proof}

With the definitions in Sections \ref{sec:data} and \ref{sec:decomp}, for any pure input state $\sigma_x^\psi = \ket{\psi}\bra{\psi}$, the global channel $\mathcal{M}$ can be decomposed into
\begin{equation}
\begin{split}
    \mathcal{M}\left(\sigma_x^\psi\right) &= \mathcal{M}\left(\frac{P_{x}^\psi\rho_x P_{x}^\psi}{\tr(\rho_x P_{x}^\psi)}\right) = \mathcal{M}\left(\frac{P_{x}^\psi\tr_C(\rho) P_{x}^\psi}{\tr(\rho_x P_{x}^\psi)}\right)\\
    &=\frac{1}{\tr(\rho_x P_x^\psi)}\sum_i\mathcal{M}(P_x^\psi \tr_C( P_{C_i} \rho P_{C_i} ) P_x^\psi)\\
    &=\frac{1}{\tr(\rho_x P_x^\psi)}\sum_i\mathcal{M}(\tr_C(P_x^\psi \rho_{x|C_i} P_x^\psi)) \tr(\rho P_{C_i})\\
    &=\frac{1}{\tr(\rho_x P_x^\psi)}\sum_i\mathcal{M}_i(\sigma_x) \tr(\rho_{x|C_i} P_{x}^\psi) p_{C_i}\\
    &=\sum_i\mathcal{M}_i(\sigma_x) \frac{\mathcal{D}_i(\sigma_x) p_{C_i}}{\sum_j \mathcal{D}_j(\sigma_x) p_{C_j}},
\end{split}
\end{equation}
where the second line utilizes the fact that $\rho$ is diagonal in $\ket{C_i}$: $\rho = \sum_i P_{C_i} \rho P_{C_i}$, and the last equality follows from 
\begin{equation}
\begin{split}
    &\tr(\rho_x P_x^\psi) = \tr(\rho P_x^\psi)=\sum_j \tr(P_{C_j}\rho P_{C_j} P_x^\psi)\\
    &= \sum_j \tr(\rho_{x|C_j} P_x^\psi)\tr(\rho P_{C_j})=\sum_j \mathcal{D}_j(\sigma_x)p_{C_j}
\end{split}
\end{equation}
As for mixed states, we note that they can always be decomposed into a linear combination of pure states. Thus following from the linearity of quantum channels, the formula for $\mathcal{M}$ acting on mixed states follows from direct linear superposition.  This completes the proof of Theorem \ref{thm:1}. 

\section{A Proposal of Quantum Generative Learning with \textit{qFedInf}} \label{app:generative}

{
We mentioned in the main text that the proposed framework \textit{qFedInf} may be applied to machine learning tasks beyond classification. Here we provide a specific example of performing quantum generative learning \cite{gao2018quantum, lloyd2018quantum, liu2018differentiable} with \textit{qFedInf}. This only serves as a preliminary proposal and we leave the detailed study of its performance and implications to future works.
}

{
In quantum generative learning, we aim to learn a generative model that can reconstruct some target quantum state $\rho_x$. In a federated learning context, each client $C_i$ only has access to a small proportion of the total data, which statistically forms a quantum state $\rho_x^{C_i}$. Thus the whole target state can be written as $\rho_x = \sum_{i} p_{C_i} \rho_x^{C_i}$, where $p_{C_i}$ is the proportion of data accessible to client $C_i$. The notations here are the same as in Section \ref{sec:data}.
}

{
We take the quantum generative adversarial network (qGAN) \cite{lloyd2018quantum} as our quantum channel $\mathcal{M}$ to perform the learning task. It's a quantum circuit that takes some fixed initial state, for example $\ket{0\cdots0}$, as its input, and outputs a quantum state, which is parameterized by the circuit parameters. Adversarial learning strategies are applied to train the circuit and the output state after training is expected to approximate the target state. Here we omit the training details as our focus is on the federated learning aspect.
}

{
In the \textit{qFedInf} framework, each client trains its own qGAN, denoted as the local channel $\mathcal{M}_i$, with its own data $\rho_x^{C_i}$. After training, we expect $\mathcal{M}_i(\ket{0\cdots0}\bra{0\cdots0})\approx\rho_x^{C_i}$. As for the density estimation part, we note that the input states are fixed to be $\ket{0\cdots0}$, so the density estimators become trivial, i.e. $\mathcal{D}_i(\ket{0\cdots0})=1$. Plug these into Equation (\ref{eqn:decomp}) and we arrive at 
\begin{equation}
    \mathcal{M}(\ket{0\cdots0}\bra{0\cdots0}) \approx \sum_i p_{C_i} \rho_x^{C_i} = \rho_x,
\end{equation}
which is exactly our goal. This is a concrete proposal of quantum federated generative learning which has not appeared in the literature so far.
}

\end{widetext}

\section*{Declarations}
\textbf{Competing interests} The author declares no competing interests.

\textbf{Funding} This work is financially supported by Zhili College, Tsinghua University.

\textbf{Availability of data and materials} All the data and materials used in this work can be accessed at \url{https://github.com/JasonZHM/quantum-fed-infer}.


\bibliography{apssamp}

\end{document}